\documentclass[twocolumn,preprintnumbers,
amsmath,amssymb,aps,prd,nofootinbib,superscriptaddress,
eqsecnum]{revtex4}
\usepackage{graphicx,color}

\makeatletter
\renewcommand{\p@subsection}{}
\makeatother

\begin{document}

\title{Anomaly-induced Chiral Mixing in Cold and Dense Matter}

\author{Chihiro Sasaki}
\affiliation{%
Institute of Theoretical Physics, University of Wroclaw,
PL-50204 Wroclaw,
Poland
}

\date{\today}

\begin{abstract}
We construct the spectral functions for light vector mesons at finite density and temperature in the presence of
a novel mixing between parity partners, induced by baryon density via the Wess-Zumino-Witten action.
As the main origin of in-medium broadening, a set of baryon resonances that strongly couple to the vector mesons
and the modifications of kaons and anti-kaons due to the Kaplan-Nelson term are included.
It is shown that the vector spectra, even with the broadening effects, exhibit sizable signatures of chiral symmetry
restoration thanks to the chiral mixing depending on three-momenta carried by the vector mesons.
Those spectral functions are used to calculate the integrated production rates of lepton pairs, and a proper binning in momenta and
potential decrease in the vector-meson masses due to chiral symmetry restoration are discussed in quantifying the signatures.
\end{abstract}

\maketitle

%%%%%%%%%%%%%%%%%%%%%%%%%%%%%%%%%%%%%%%%%%%%%%%%%%%%
\section{Introduction}
%%%%%%%%%%%%%%%%%%%%%%%%%%%%%%%%%%%%%%%%%%%%%%%%%%%

Light vector mesons, especially their dynamical properties arising from spontaneously broken chiral symmetry in QCD,
have been studied extensively in various approaches~\cite{Rapp,HH,RWvH,Rapp:2011zz,FS}.
In-medium modifications of vector spectral functions were anticipated due to the interactions with a hadronic medium,
which carry imprints of restoration of chiral symmetry at high temperature and/or baryon density.
Strong modifications have been indeed measured via dilepton production in heavy-ion collisions~\cite{NA60,vHR},
whereas it remains inconclusive how to quantify those modifications of vector mesons as the direct consequence of
chiral symmetry restoration since the vector spectra themselves are not the order parameter.

The ideal way is to measure the spectral functions both in the vector and its parity partner, axial-vector channels
to examine if the two spectra are nearly degenerate as expected with unbroken chiral symmetry.
Given the difficulty that it is elusive to construct the axial-vector spectrum in heavy-ion experiments,
the key phenomenon is that the vector mesons mix with the axial-vector mesons at finite temperature and density,
known as chiral mixing.
The chiral mixing effect induced by pions can be quantified in a model-independent way at low temperature/density~\cite{DEI,Krippa98,Chanfray}.
Its extrapolation toward the chiral symmetry restoration requires a caution.
There exists a systematic calculation~\cite{HRhot} to construct the in-medium axial-vector spectral function at finite temperature
from the vector correlator that successfully describes the dilepton data, via the Weinberg sum rules~\cite{WSR}.
It was reported there that the $a_1$ meson mass drops substantially toward the $\rho$ mass together with a width broadening.
The observed trend is consistent with another study on the chiral mixing in which the Nambu-Goldstone bosons and vector mesons with
opposite parity were explicitly included in a chiral Lagrangian to perform a systematic computation at loop level~\cite{HSWhot}.
It was shown that the chiral symmetry restoration forces the chiral mixing to vanish, which ensures the vector and axial-vector
spectra degenerate while it is totally distinct from the result of a naive extrapolation of the mixing theorem leading to a maximal mixing~\cite{UBW}.

There exists a novel class of chiral mixing induced by baryon density via anomaly in a dense medium.
It was first recognized in a holographic QCD with the Chern-Simons action and the emergent mixing operator results in
the modified dispersion relations of transverse polarizations of vector and axial-vector states~\cite{DHholo}.
The low-energy Lagrangian was then reformulated in the standard 4-dimensional chiral effective theory and
a brief estimate of the mixing strength as a function of baryon density was given in the mean field approximation~\cite{HSdense}.
Although the resultant mixing is smaller than the holographic estimate by one order of magnitude,
the chiral mixing is reinforced with decreasing mass difference between parity partners, i.e. the onset of chiral symmetry
restoration, and leads to the structural change of the vector spectra and the dilepton production rates~\cite{Sasaki}.

We are yet to extend the previous calculations of vector spectra in the presence of density-induced chiral mixing.
The mixing effect results in the spectral function somewhat broadened since the transverse modes obey the non-standard dispersion
relations leading to a downward shift of their masses and the final spectrum as a superposition of all the polarization states
is no longer in the Breit-Wigner distribution~\cite{HSdense,Sasaki} where the main origin of standard broadening due to baryon resonances
that strongly couple to the vector mesons in nuclear matter was not included.

In this paper, we will introduce those baryon resonances into the computation of in-medium vector spectral functions for the
$\rho/\omega$ states at finite density and low temperature. We will also include modifications of kaons and anti-kaons
in nuclear matter due to the Kaplan-Nelson term~\cite{kaon} to extend our previous calculation of the $\phi$ spectral function.
The updated set of spectral functions will be used to calculate the production rates of dilepton to examine to what extent
the signatures of chiral symmetry restoration via the chiral mixing would stay in such observable quantities in heavy-ion collisions.

%%%%%%%%%%%%%%%%%%%%%%%%%%%%%%%%%%%%%%%%%%%%%%%%%%%%
\section{Chiral mixing via the Wess-Zumino-Witten term}
%%%%%%%%%%%%%%%%%%%%%%%%%%%%%%%%%%%%%%%%%%%%%%%%%%%

Non-vanishing baryon chemical potential violates explicitly charge-conjugation invariance.
Consequently, effective Lagrangians contain a set of new operators that are prohibited
in matter-free space. In the system composed of the Nambu-Goldstone bosons as well as the lowest vector
$V_\mu$ and axial-vector $A_\mu$ mesons, there exists an operator leading to a direct mixing of
vector with axial-vector states, the chiral mixing, in the form of
\begin{equation}
{\mathcal L}_{\rm mix} = 2c\,\epsilon^{0\mu\nu\lambda}\mbox{tr}\left[
\partial_\mu V_\nu\cdot A_\lambda + \partial_\mu A_\nu\cdot V_\lambda
\right]\,,
\label{lag1}
\end{equation}
with the total anti-symmetric tensor $\epsilon^{0123}=1$ and a parameter $c$ to be fixed later.
This term can be deduced from the Chern-Simons action in a model based on AdS/CFT correspondence~\cite{DHholo},
and alternatively from the Wess-Zumino-Witten (WZW) action~\cite{KM}.
The mixing term (\ref{lag1}) modifies dispersion relations for the transverse polarizations as
\begin{equation}
p_0^2 - \vec{p}^2
= \frac{1}{2}\left[
m_V^2 + m_A^2 \pm \sqrt{(m_A^2 - m_V^2)^2 + 16 c^2 \vec{p}^2}
\right]\,,
\label{disp}
\end{equation}
where the lower sign is for the vector and the upper one for the axial-vector mesons.
The longitudinal polarizations obey the standard dispersion relation, $p_0^2 - \vec{p}^2 = m_{V,A}^2$.

The holographic model~\cite{DHholo} yields a rather strong density dependence in the mixing parameter,
$c \simeq 1$ GeV at the normal nuclear matter density $\rho_0$. This, however, results in the onset of
vector meson condensation only slightly above $\rho_0$, which is an apparent drawback of the large $N_c$
prescription that the approaches with AdS/CFT depend on.
In contrast, the standard chiral approach with the gauged WZW action~\cite{HSdense} yields a moderate
density-effect, with the expectation value of iso-scalar vector meson $\langle\omega_0\rangle = g_{\omega NN}\cdot\rho_B/m_\omega^2$,
leading to $c = g_{\omega\rho a_1}\langle\omega_0\rangle \simeq 0.1$ GeV at $\rho_0$.

Whereas the weak mixing is negligible in the propagators, thus in vector spectral functions as well,
the key feature is that the mixing strength linearly increases with baryon density $\rho_B$.
In Ref.~\cite{Sasaki}, it has been shown that the restoration of chiral symmetry enhances the chiral mixing
and the structural changes in vector spectra serve as a clear signature of chiral restoration in cold and
dense matter. Such a drastic deformation further yields a non-negligible contribution to dilepton production rates,
and thus in principle it is measurable in heavy-ion collisions at FAIR, NICA and J-PARC.
Our previous calculations~\cite{Sasaki} were performed without explicit baryonic resonances arising from
the direct coupling of vector mesons to the surrounding nucleons. Those resonances are known to generate significant
modifications in self-energies in the $\rho/\omega$ sectors.
In the invariant-mass region relevant to the $\phi$ meson, there are no $\phi N$ resonances but the kaon and anti-kaon
change their properties in nuclear matter~\cite{kaon}.
In the next section, we will introduce those characteristic effects to estimate the interplay between the vector mesons
and the chiral mixing induced by density.

%%%%%%%%%%%%%%%%%%%%%%%%%%%%%%%%%%%%%%%%%%%%%%%%%%%%
\section{Vector spectral functions}
%%%%%%%%%%%%%%%%%%%%%%%%%%%%%%%%%%%%%%%%%%%%%%%%%%%

We begin with the current-current correlation functions in matter;
\begin{equation}
G_{V,A}^{\mu\nu}(p_0,\vec{p})
= P_L^{\mu\nu}G_{V,A}^L(p_0,\vec{p}) + P_T^{\mu\nu}G_{V,A}^T(p_0,\vec{p})\,,
\end{equation}
with the polarization tensors
\begin{eqnarray}
P_{T,\mu\nu}
&=&
g_{\mu i}\left(\delta_{ij}-\frac{\vec{p}_i\vec{p}_j}{\vec{p}^2}\right)g_{j\nu}\,,
\nonumber\\
P_{L,\mu\nu}
&=&
-\left(g_{\mu\nu}-\frac{p_\mu p_\nu}{p^2}\right) - P_{T,\mu\nu}\,.
\end{eqnarray}
Formulating the chiral Lagrangian for pions, vector and axial-vector mesons in the non-linear realization~\cite{BKY,HSghls},
the longitudinal and transverse parts read~\cite{HSdense}
\begin{eqnarray}
&&
G_V^L = \left(\frac{g_V}{m_V}\right)^2\frac{-s}{D_V^L}\,,
\quad
G_V^T = \left(\frac{g_V}{m_V}\right)^2\frac{-sD_A^T + 4c^2\vec{p}^2}{D_V^TD_A^T - 4c^2\vec{p}^2}\,,
\nonumber\\
&&
G_A^L = \left(\frac{g_A}{m_A}\right)^2\frac{-s}{D_A^L}\,,
\quad
G_A^T = \left(\frac{g_A}{m_A}\right)^2\frac{-sD_V^T + 4c^2\vec{p}^2}{D_V^TD_A^T - 4c^2\vec{p}^2}\,,
\nonumber\\
\label{spectra}
\end{eqnarray}
with $s = p_0^2 - \vec{p}^2$ and the coupling of the vector/axial-vector meson to the
corresponding current $g_{V,A}$ as well as the propagator inverse without the mixing
$D_{V,A}^{L,T} = s - m_{V,A}^2 - \Sigma_{V,A}^{L,T}(s)$.
The self-energies $\Sigma_V^{L,T}$ will be computed in the presence of baryon resonances
in the $\rho$-meson sector and the modified kaons in the $\phi$-meson sector.
The spin-averaged correlators are given by $G_{V,A} = \frac{1}{3}\left(G_{V,A}^L + 2G_{V,A}^T\right)$.
The labels $(V,A)$ refer to the iso-vector $(\rho,a_1)$ and
the iso-singlet $(\omega,f_1(1285))$ and $(\phi,f_1(1420))$ mesons.
The coupling constants to the vector current are related to $g_\rho = 0.119$ GeV$^2$
via chiral symmetry, so that
\begin{equation}
g_\omega = \frac{1}{3}\frac{m_\omega^2}{m_\rho^2}g_\rho\,,
\quad
g_\phi = \frac{\sqrt{2}}{3}\frac{m_\phi^2}{m_\rho^2}g_\rho\,.
\end{equation}

We shall focus on the vector spectral functions at chiral symmetry restoration since
the mixing effect becomes maximal there. We will also neglect meson-loop effects which are
small at low temperature, $T \ll m_\pi$, relevant to our study.
The in-medium masses are nearly degenerate, $m_V \simeq m_A$, leading to $G_V \simeq G_A$.
As demonstrated in Ref.~\cite{Sasaki}, the central quantities,
i.e. the mass difference between vector and axial-vector mesons and the mixing strength, were estimated
at temperature $T = 50$ MeV, and found to be $\delta m = 47$ MeV and $c = 98$ MeV with the chiral crossover
$\rho_B^{\rm crit}/\rho_0 = 2.5$, within an extended parity-doublet model that describes not only the properties of
nuclear ground states in symmetric matter but also those of neutron star matter~\cite{Marczenko}.
Below, we will assume for simplicity the vector-meson mass being independent of density and temperature.

%%%%%%%%%%%%%%%%%%%%%%%%%%%%%%%%%%%%%
\subsection{$\rho$/$\omega$ mesons}
%%%%%%%%%%%%%%%%%%%%%%%%%%%%%%%%%%%%%

The importance of $\rho N$ interactions in nuclear matter was first discussed in Ref.~\cite{FP}
where two p-wave states, $N(1720)$ and $\Delta(1905)$, were considered because of their large branching
ratios into $\rho$ and $N$. Due to the $\rho N$ interactions, the expected three states, the $\rho$ meson
and two resonance-hole states, are all mixed, and lead to the spectral function strongly smeared.

Since the photo-absorption data requires the inclusion of additional low-lying resonances, we will include
a larger set of resonances following Ref.~\cite{RUBW}, summarized in Table~\ref{res}.
%%%%%%%%%%%%%%%%%%%%%%%%%%%%%%%%%%%%%%%%%%%%%%%%%%%%%%%%%%%%%%%
\begin{table}
\begin{center}
\begin{tabular}{@{\extracolsep{\fill}}cccccc}
\hline
B & $l_{\rho N}$ & SI & $\Gamma^0_{\rho N}$ [MeV] & $\left({f^2_{\rho BN}}/{4\pi}\right)^2$ & $\Gamma^{\rm med}$ [MeV] \\
\hline
$\Delta(1232)$ & $p$ & $16/9$ & $-$ & $23.2$ & $15$ \\
$N(1520)$ & $s$ & $8/3$ & $24$ & $5.5$ & $250$\\
$\Delta(1620)$ & $s$ & $8/3$ & $22.5$ & $0.7$ & 50\\
$\Delta(1700)$ & $s$ & $16/9$ & $45$ & $1.2$ & $50$\\
$N(1720)$ & $p$ & $8/3$ & $105$ & $9.2$ & $50$\\
$\Delta(1905)$ & $p$ & $4/5$ & $210$ & $18.5$ & $50$\\
\hline
\end{tabular}
\end{center}
\caption{
Set of baryon resonances and their properties~\cite{RUBW}.
}
\label{res}
\end{table}
%%%%%%%%%%%%%%%%%%%%%%%%%%%%%%%%%%%%%%%%%%%%%%%%%%%%%%%%%%%%%%%%%%%
Employing the interaction Lagrangians for those resonances given in~\cite{FP}, the in-medium self-energy
of transverse and longitudinal $\rho$ mesons, to the lowest order in density, read
\begin{eqnarray}
\Sigma_\rho^T(p_0,\vec{p})
&=&
\sum_{\rm p-wave}
SI\,\frac{f^2_{\rho BN}}{m_\rho^2}\,F(\vec{p}^2)\,\vec{p}^2\,\rho_B\,
\frac{(E_p^B - m_N)}{p_0^2 - (E_p^B - m_N)^2}
\nonumber\\
&+&
\sum_{\rm s-wave}
SI\,\frac{f^2_{\rho BN}}{m_\rho^2}\,F(\vec{p}^2)\,p_0^2\,\rho_B\,
\frac{(E_p^B - m_N)}{p_0^2 - (E_p^B - m_N)^2}\,,
\nonumber\\
\Sigma_\rho^L(p_0,\vec{p})
&=&
\sum_{\rm s-wave}
SI\,\frac{f^2_{\rho BN}}{m_\rho^2}\,F(\vec{p}^2)\,p_0^2\,\rho_B\,
\frac{(E_p^B - m_N)}{p_0^2 - (E_p^B - m_N)^2}\,,
\nonumber\\
\end{eqnarray}
where
\begin{eqnarray}
E_p^B(p_0,\vec{p})
&=&
\sqrt{\vec{p}^2 + m_B} - \frac{i}{2}\Gamma_B(p_0,\vec{p})\,,
\nonumber\\
\Gamma_B(p_0,\vec{p})
&=&
\Gamma_{B\to \rho N}^0(p_0,\vec{p}) + \Gamma_B^{\rm med}\,\frac{\rho_B}{\rho_0}\,,
\end{eqnarray}
and a form factor $F(\vec{p}^2) = \Lambda^2/(\Lambda^2+\vec{p}^2)$
with $\Lambda = 600$ MeV~\cite{RUBW}.
The spin-isospin factors $SI$ and the parameters $f_{\rho BN}$ for the resonances included are summarized in
Table~\ref{res}.

The $\Gamma_{B\to\rho N}^0$ in the above equation represents the full width modified by the phase space for resonances.
We will follow the prescription given in Ref.~\cite{FP} to introduce the energy and momentum dependence
in the static approximation for the nucleon;
\begin{equation}
\Gamma_{B\to\rho N}^0(p_0,\vec{p})
=
\Gamma_{B\to\rho N}^0\,\left(\frac{p}{p_B}\right)^3\,,
\end{equation}
where
\begin{eqnarray}
p
&=&
\sqrt{\left[(\bar{s} - m_N^2 - m_\pi^2)^2 - 4m_N^2 m_\pi^2\right]/4\bar{s}}\,,
\nonumber
\\
p_B
&=&
\sqrt{\left[(m_B^2 - m_N^2 - m_\pi^2)^2 - 4m_N^2 m_\pi^2\right]/4m_B^2}\,,
\end{eqnarray}
with $\bar{s} = (p_0 + m_N)^2 - \vec{p}^2$.
The final self-energy of $\rho$ meson reads $\Sigma_{\rho,{\rm tot}}^{T,L} = -im_\rho\Gamma_\rho + \Sigma_\rho^{T,L}$
with the vacuum $\rho$-meson width $\Gamma_\rho(s)$.

The $\omega$ meson in nuclear matter, similarly to the $\rho$ meson, gets broader but with little mass shift,
and its spectral function still exhibits a distinct maximum~\cite{Rapp,RWvH,Rapp:2011zz}.
Therefore, we shall assume for simplicity that the in-medium width at chiral restoration is larger than
its vacuum value by a constant factor, $7$, reproducing the value $\sim 60$ MeV at $\rho_B/\rho_0 = 2$
found in a coupled channel approach implementing $\omega N$ resonances~\cite{giessen}.

\begin{figure*}
\includegraphics[width=5.6cm]{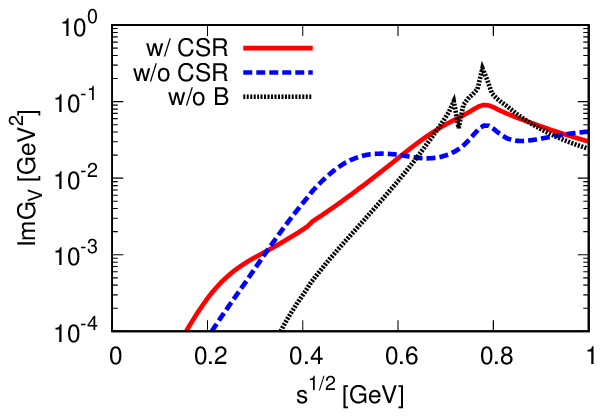}
\includegraphics[width=5.6cm]{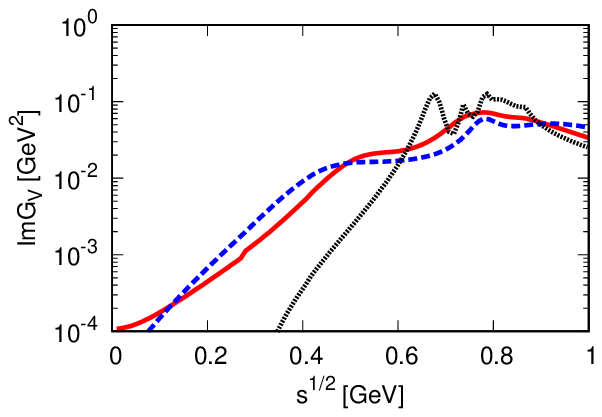}
\includegraphics[width=5.6cm]{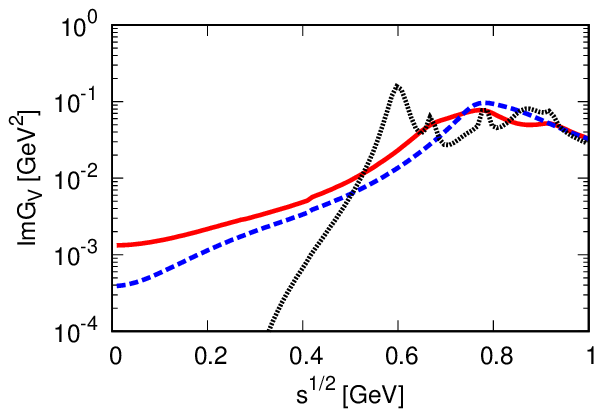}
\caption{The $\rho/\omega$ spectral function at chiral crossover $\rho_B^{\rm crit}/\rho_0 = 2.5$ and $T=50$ MeV
with finite three-momenta, $\vec{p}=0.1, 0.5, 1.0$ GeV, from left to right.
The curves without chiral symmetry restoration (CSR) were computed such that the mass difference was kept to be
its vacuum value, i.e. $\delta m = 0.49$ GeV at any $\rho_B$ and $T$.
The curves "w/o B" represent the results without baryon resonances, taken from our earlier study in Ref.~\cite{Sasaki}.
}
\label{sf_rho}
\end{figure*}

Fig.~\ref{sf_rho} represents the spectral function of $\rho/\omega$ states at the chiral crossover at $T=50$ MeV
with selected three-momenta, $\vec{p}=0.1, 0.5, 1.0$ GeV.
%%%%%%%%%%%%%%%%%%%%%%%%%%%%%%%%%%%%%%%%%%%%%%%%%%%%%%%%%
%%%%%%%%%%%%%%%%%%%%%%%%%%%%%%%%%%%%%%%%%%%%%%%%%%%%%%%%%%%%
The three s-wave states in Table~\ref{res} carry negative parity, and they are expected to decrease their masses
toward the corresponding parity partners near the chiral symmetry restoration (CSR),
just as the lowest-lying baryons do~\cite{fastsum}.
We thus assume that the masses of the three resonances with negative parity are reduced by a similar amount, 25\%.

For the smallest momentum $\vec{p}=0.1$ GeV, the mixing effect has little importance. The difference between the two
lines is due to the s-wave baryonic resonances whose masses are reduced when CSR is realized in dense matter.
With increasing $\vec{p}$, the p-wave states come into the spectral function. The spectrum with $\vec{p}=1$ GeV
is an admixture of shifted transverse and unshifted longitudinal polarization states of vector and axial-vector mesons
via chiral mixing, on top of the in-medium broadening induced dominantly by the p-wave resonances.
When CSR is not imposed, the former effect, the chiral mixing, is negligible so that the spectrum is broadened
just due to the latter effect.
The importance of including baryon resonances is evident: the emergent bumpy structure due to CSR found in our
earlier study \cite{Sasaki} is smeared.

\begin{figure*}
\includegraphics[width=8.5cm]{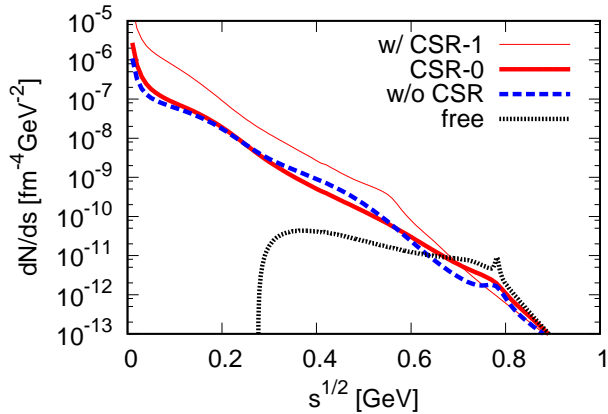}
\caption{The dilepton production rates at chiral crossover $\rho_B^{\rm crit}/\rho_0 = 2.5$ and $T=50$ MeV,
integrated over three-momenta in the rage of $0 \leq |\vec{p}| \leq 2$ GeV.
The curve labeled "CSR-0" was calculated with the mass of vector meson being the same value as in matter-free space,
whereas the curve labeled "CSR-1" was obtained by implementing a reduction of vector meson mass to $550$ MeV as suggested in Ref.~\cite{Kim:2021xyp}}
\label{dl_rho}
\end{figure*}

The production rate of a lepton pair emitted from dense matter via a virtual photon is readily calculated.
The differential rate at finite $T$ and $\mu_B$ is given
in terms of the imaginary part of the vector current correlator~\cite{RWvH} by
\begin{equation}
\frac{dN}{d^4p}(p_0,\vec{p};T,\mu_B)
= \frac{\alpha^2}{\pi^3 s}\frac{\mbox{Im}G_V(p_0,\vec{p};T,\mu_B)}{e^{p_0/T}-1}\,,
\end{equation}
with $\alpha=e^2/4\pi$ the electromagnetic coupling constant.
The three-momentum integrated rate reads
\begin{equation}
\frac{dN}{ds}(s;T,\mu_B)
= \int\frac{d^3\vec{p}}{2p_0}\frac{dN}{d^4p}(p_0,\vec{p};T,\mu_B)\,.
\end{equation}
In Fig.~\ref{dl_rho}, the integrated rate in the range of $0 \leq |\vec{p}| \leq 2$ GeV at chiral crossover with $T=50$ MeV is presented.
%%%%%%%%%%%%%%%%%%%%%%%%%%%%%%%%%%%%%%%%
%%%%%%%%%%%%%%%%%%%%%%%%%%%%%%%%%%%%%%%%
The result obtained from the vector spectra shown in Fig.~\ref{sf_rho}, labeled "CSR-0", exhibits a substantial broadening effect
mainly due to the p-wave baryon resonances, and consequently it does not differ much from the baseline without CSR.
We have assumed so far that the vector meson mass at chiral crossover remains its vacuum value, which can be relaxed.
In fact, the recent study using QCD sum rules applied to the vacuum with unbroken chiral symmetry suggests that the degenerate
parity partners of vector states carry the mass of $500$-$600$ MeV~\cite{Kim:2021xyp}.
Introducing such a reduction into the vector spectra results in a much larger contribution in the production rate
below $\sqrt{s} = 0.7$ GeV, labeled "CSR-1".
In practice, the signatures of CRS in the rates are to a large extent diminished by the p-wave baryon resonances and it would be
hard to measure them in heavy-ion experiments, unless the masses of vector states would decrease near the chiral symmetry restoration
by $20$-$30$ \%.

%%%%%%%%%%%%%%%%%%%%%%%%%%%%
\subsection{$\phi$ meson}
%%%%%%%%%%%%%%%%%%%%%%%%%%%%

The $\phi$ mesons predominantly decay into $K\bar{K}$ pairs, and
the dressing of kaon cloud mainly modifies the $\phi$ meson in nuclear matter.
The $SU(3)$ chiral Lagrangian in the mean field approximation gives
in-medium masses of kaon and anti-kaon as~\cite{kaon}
\begin{eqnarray}
m_K^\ast
&=&
\left[m_K^2 - a_K\rho_S + \left(b_K\rho_B\right)^2\right]^{1/2} + b_K\rho_B\,,
\nonumber\\
m_{\bar{K}}^\ast
&=&
\left[m_K^2 - a_{\bar{K}}\rho_S + \left(b_K\rho_B\right)^2\right]^{1/2} - b_K\rho_B\,,
\label{kaon_disp}
\end{eqnarray}
where $\rho_S$ represents the nuclear scalar density and the three parameters are given by
$a_K = a_{\bar{K}}=\Sigma_{KN}/f_\pi^2$ with the kaon-nucleon sigma term $\Sigma_{KN}$ and
$b_K = 3/(8 f_\pi^2)$, respectively.
Given the fact that there exist large uncertainties in $\Sigma_{KN}$ and difficulties in dealing with higher-order
corrections systematically,
the parameters $a_K$ and $a_{\bar{K}}$ have been determined from the kaon production data in
heavy-ion collisions and found $a_K = 0.22$ GeV$^2$fm$^3$ and
$a_{\bar{K}} = 0.45$ GeV$^2$fm$^3$, respectively~\cite{Li:1997tz}.
One finds the in-medium kaon masses to be $m_K^\ast = 560$ MeV and $m_{\bar{K}}^\ast = 230$ MeV
at the chiral crossover, $\rho_B^{\rm crit}/\rho_0 = 2.5$.

Accordingly, the decay width of $\phi$ meson is modified as well. We shall adopt the form~\cite{Chung:1998ev},
\begin{eqnarray}
\Gamma_\phi(s)
&=&
\frac{g^2_{\phi K\bar{K}}}{3\pi}\frac{k(s)^3}{s}\,,
\nonumber\\
k(s)
&=&
\frac{1}{2\sqrt{s}}
\left[\left(s - (m_K^\ast + m_{\bar{K}}^\ast)^2\right)
\left(s - (m_K^\ast - m_{\bar{K}}^\ast)^2\right)\right]^{1/2}\,,
\nonumber\\
\end{eqnarray}
with the coupling constant $g_{\phi K\bar{K}}^2/4\pi = 1.69$.
Assuming that the $\phi$ meson mass is little shifted in nuclear matter as studied in
the $SU(3)$ coupled channel approach~\cite{Oset:2000eg},
the kaon and anti-kaon obeying the above relations (\ref{kaon_disp}) lead to
the effective width $\Gamma_\phi(s=m_\phi^2) = 61$ MeV.

\begin{figure*}
\includegraphics[width=5.6cm]{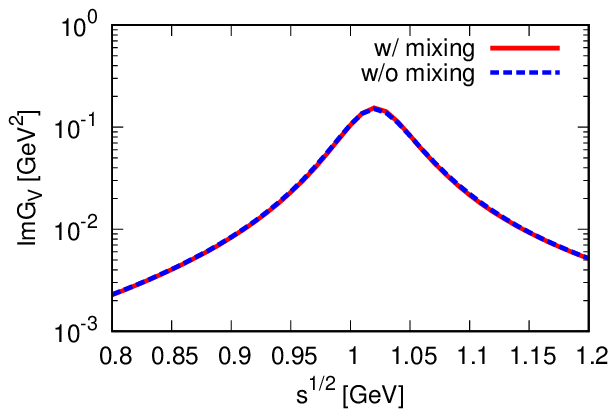}
\includegraphics[width=5.6cm]{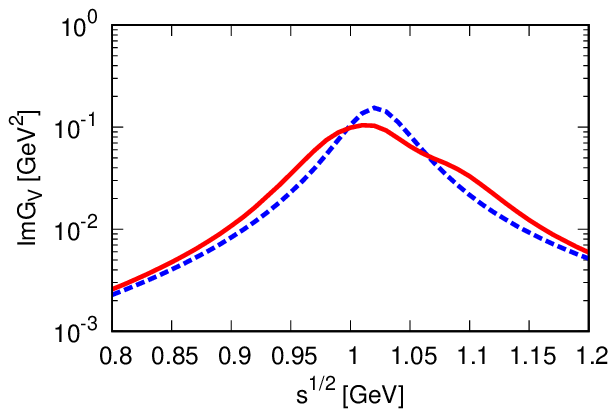}
\includegraphics[width=5.6cm]{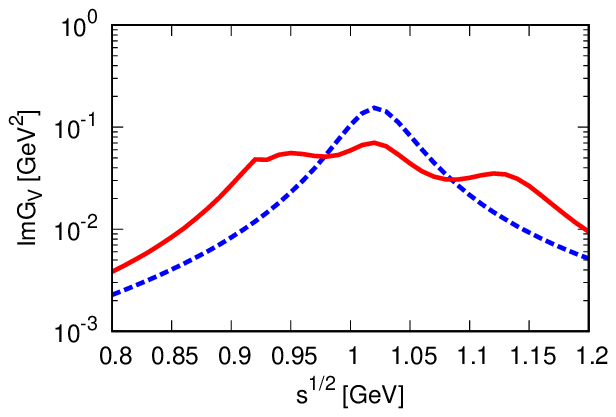}
\caption{The $\phi$ spectral function at chiral crossover $\rho_B^{\rm crit}/\rho_0 = 2.5$ and $T=50$ MeV
with finite three-momenta, $\vec{p}=0.1, 0.5, 1.0$ GeV, from left to right.}
\label{sf_phi}
\end{figure*}

Given the modified width of $\phi$ meson, the spectral function can be readily computed
at chiral crossover, for a fixed three-momentum $\vec{p}$, as displayed in Fig.~\ref{sf_phi}.
%%%%%%%%%%%%%%%%%%%%%%%%%%%%%%%%%%%%%%%%%%%%%%%%%%%%%%%
%%%%%%%%%%%%%%%%%%%%%%%%%%%%%%%%%%%%%%%%%%%%%%%%%%%%%%%
Recall that the chiral mixing effect (\ref{disp}) is always accompanied by finite $\vec{p}$.
The signatures of CSR thus become stronger for larger $\vec{p}$, whose contributions are more Boltzmann suppressed
in dilepton rates though.
Therefore, a careful binning of dilepton data in three-momenta, say $|\vec{p}| > 0.5$ GeV, would be crucial
to extract the signals of restored chiral symmetry.

\begin{figure*}
\includegraphics[width=5.6cm]{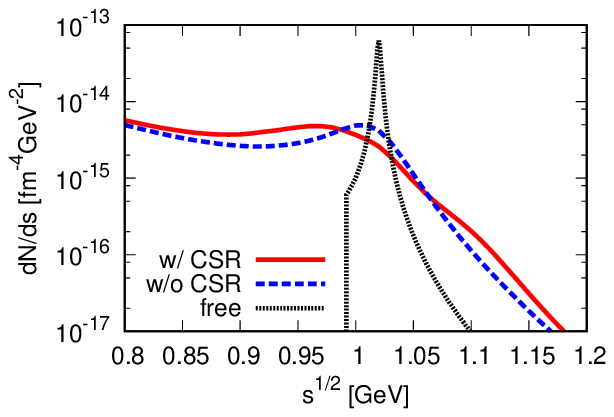}
\includegraphics[width=5.6cm]{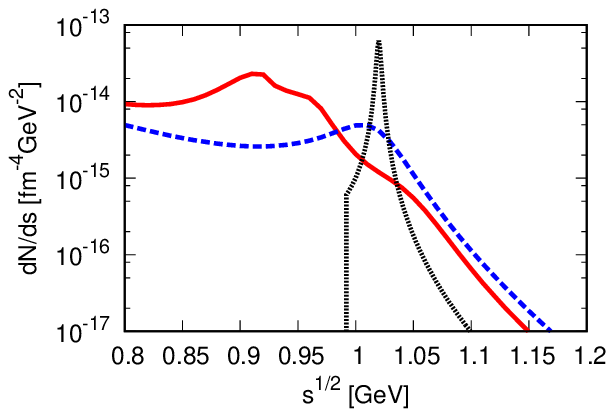}
\includegraphics[width=5.6cm]{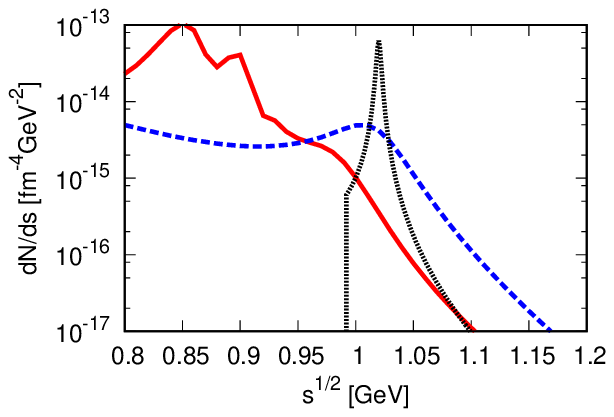}
\caption{The dilepton production rates at chiral crossover $\rho_B^{\rm crit}/\rho_0 = 2.5$ and $T=50$ MeV,
integrated over three-momenta in the rage of $0.5 \leq |\vec{p}| \leq 2$ GeV.
The $\phi$ meson mass was assumed to be the same as in vacuum (left), reduced by $5$\% (middle) and $10$\% (right).}
\label{dl_phi}
\end{figure*}

The corresponding production rates of lepton pairs integrated over $\vec{p}$ above $0.5$ GeV up to $2$ GeV
are shown in Fig.~\ref{dl_phi}.
%%%%%%%%%%%%%%%%%%%%%%%%%%%%%%%%%%%%%
%%%%%%%%%%%%%%%%%%%%%%%%%%%%%%%%%%%%%
The clear peak due to the chiral mixing seen at $\sqrt{s}\simeq 1.1$ GeV \cite{Sasaki} where no modifications of kaons were included
is now smeared because of in-medium broadening.
When the mass of $\phi$ meson decreases toward CSR, even by a few percent, the mixing effect from the $\phi$ and its counterpart
becomes enhanced and generates a sizable difference from the result without CSR.

\begin{figure*}
\includegraphics[width=8.5cm]{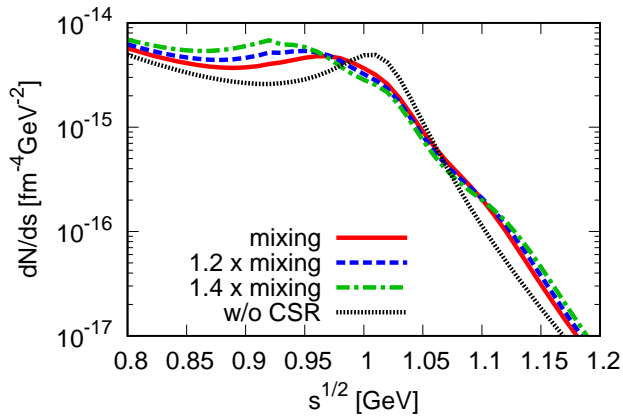}
\caption{The same as in Fig.~\ref{dl_phi} (left), but for various strengths of chiral mixing.}
\label{large_mixing}
\end{figure*}

So far, our calculations were performed with the mixing strength estimated in the mean field approximation that also
yields a specific density of chiral crossover, $2.5\,\rho_0$~\cite{Sasaki}.
Since the chiral mixing is linearly proportional to the net baryon density $\rho_B$, one can accommodate potential options,
e.g. a stronger mixing and/or higher critical density of chiral symmetry restoration, into our computation by tuning the mixing
strength.
In Fig.~\ref{large_mixing}, we show the dilepton rates at $T=50$ MeV for several values of the mixing parameter in dense matter.
%%%%%%%%%%%%%%%%%%%%%%%%%%%%%%%%%%
%%%%%%%%%%%%%%%%%%%%%%%%%%%%%%%%%%%
The curve labeled "$1.2\,\times$ mixing" represents the result at the same chiral crossover $\rho_B^{\rm crit}=2.5\,\rho_0$
with a stronger mixing by $20$\%, or alternatively at a higher crossover density $3\,\rho_0$ with the same mixing strength.
Similarly, the curve labeled "$1.4\,\times$ mixing" depicts two scenarios, i.e. a stronger mixing by $40$\% alternating with
a higher critical density $3.5\,\rho_0$.
The signatures of CSR become stronger with larger baryon density leading to either stronger mixing or higher critical density, or even both.
To better quantify the in-medium effects and their consequences on dilepton production,
a more refined approach beyond the mean field approximation would be required.

%%%%%%%%%%%%%%%%%%%%%%%%%%%%%%%%%%%%%%%%%%%%%%%%%%%%
\section{Conclusions}
%%%%%%%%%%%%%%%%%%%%%%%%%%%%%%%%%%%%%%%%%%%%%%%%%%%

We have carried out the computation of in-medium spectral functions for the light vector mesons in the
presence of novel chiral mixing induced by density, by implementing relevant baryon resonances and modifications of
kaons and anti-kaons which yield significant broadening of those spectra in dense nuclear matter.
The emergence of chiral symmetry restoration (CSR) can be found maximally via the chiral mixing~\cite{Sasaki},
and this property stays with the baryon resonances and modified kaons.

The $\rho/\omega$ spectra at chiral crossover exhibit distinct features from those without CSR in a wide range of
three momenta $\vec{p}$;
For small $\vec{p}$, the mixing effect is negligible but the s-wave resonances, which carry negative parity, change
their masses because of CSR and generate a clear difference from the scenario without CSR.
For large $\vec{p}$, the p-wave states play the major role to broaden the spectra, whereas the chiral mixing effect becomes
enlarged as well.
The $\phi$ spectra at chiral crossover are not altered much compared with the $\rho/\omega$, but those carrying higher $\vec{p}$
show stronger signatures of CSR as naturally expected from the modified dispersion relations due to the chiral mixing.

The observable quantities, such as dilepton production rates, are rather insensitive to CSR since they are integrated over $\vec{p}$.
This is particularly the case in the $\rho/\omega$ sector where the signatures of CSR are to a large extent diminished,
unless the vector meson masses would decrease by a few hundred MeV as suggested in the recent analysis based on QCD sum rules~\cite{Kim:2021xyp}.
The rates from $\phi$ mesons also show a strong sensitivity to the $\phi$ meson mass at chiral crossover, even with a slight decrease
by a few percent leading to a sizable difference from the "no CSR" scenario around the invariance mass of $1$ GeV.

A higher crossover density and/or stronger chiral mixing are regarded as theoretical options in dense nuclear matter.
If either one would be realized, the signals of CSR both in the vector spectra and dilepton rates become stronger,
although the quantitative estimate will rely on a precise value of the chiral mixing that requires a more refined prescription
beyond the approximations employed in this study.

Chiral spin symmetry has been recently shown to emerge in a temperature range of $T/T_{\rm ch} = 2-3$ with
chiral crossover temperature $T_{\rm ch}$~\cite{Glozman:2022lda}, in which the new symmetry manifests itself
in the nucleon parity doublet.
Its phenomenological consequences via parity doubling will be explored in the domain of dense QCD.

%%%%%%%%%%%%%%%%%%%%%%%%%%%%%%%%%%%%%%%%%%%%%%%%%%%%%%%%
\subsection*{Acknowledgments}
%%%%%%%%%%%%%%%%%%%%%%%%%%%%%%%%%%%%%%%%%%%%%%%%%%%%%%

The author acknowledges stimulating discussions with
T.~Galatyuk and K.~Redlich.
This work has been partly supported
by the Polish Science Foundation (NCN) under
OPUS Grant No. 2018/31/B/ST2/01663.

%%%%%%%%%%%%%%%%%%%%%%%%%%%%%%%%%%%%%%%%%%%%%%%%%%%%%%%%
%%%%%%%%%%%%%%%%%%%%%%%%%%%%%%%%%%%%%%%%%%%%%%%%%%%%%%%%%


\begin{thebibliography}{50}
\bibitem{Rapp}
  R.~Rapp,
  %``Signatures of thermal dilepton radiation at RHIC,''
  Phys.\ Rev.\ C {\bf 63}, 054907 (2001).

\bibitem{HH}
  R.~S.~Hayano and T.~Hatsuda,
  %``Hadron properties in the nuclear medium,''
  Rev.\ Mod.\ Phys.\  {\bf 82}, 2949 (2010).

\bibitem{RWvH}
  R.~Rapp, J.~Wambach and H.~van Hees,
  %``The Chiral Restoration Transition of QCD and Low Mass Dileptons,''
  Landolt-Bornstein {\bf 23}, 134 (2010).

\bibitem{Rapp:2011zz}
R.~Rapp, B.~Kampfer, A.~Andronic, D.~Blaschke, C.~Fuchs, M.~Harada, T.~Hilger, M.~Kitazawa, T.~Kunihiro and P.~Petreczky, \textit{et al.}
%``In-medium excitations,''
Lect. Notes Phys. \textbf{814}, 335-529 (2011).

\bibitem{FS}
  K.~Fukushima and C.~Sasaki,
  %``The phase diagram of nuclear and quark matter at high baryon density,''
  Prog.\ Part.\ Nucl.\ Phys.\  {\bf 72}, 99 (2013).

\bibitem{NA60}
  R.~Arnaldi {\it et al.} [NA60 Collaboration],
  %``First measurement of the rho spectral function in high-energy nuclear collisions,''
  Phys.\ Rev.\ Lett.\  {\bf 96}, 162302 (2006).

\bibitem{vHR}
  H.~van Hees and R.~Rapp,
  %``Comprehensive interpretation of thermal dileptons at the SPS,''
  Phys.\ Rev.\ Lett.\  {\bf 97}, 102301 (2006).

\bibitem{DEI}
  M.~Dey, V.~L.~Eletsky and B.~L.~Ioffe,
  %``Mixing of vector and axial mesons at finite temperature: an Indication towards chiral symmetry restoration,''
  Phys.\ Lett.\ B {\bf 252}, 620 (1990).

\bibitem{Krippa98}
  B.~Krippa,
  %``Chiral symmetry and mixing of axial and vector correlators in matter,''
  Phys.\ Lett.\ B {\bf 427}, 13 (1998).

\bibitem{Chanfray}
  G.~Chanfray, J.~Delorme and M.~Ericson,
  %``Chiral symmetry restoration and parity mixing,''
  Nucl.\ Phys.\ A {\bf 637}, 421 (1998).

\bibitem{HRhot}
  P.~M.~Hohler and R.~Rapp,
  %``Is $\rho$-Meson Melting Compatible with Chiral Restoration?,''
  Phys.\ Lett.\ B {\bf 731}, 103 (2014).

\bibitem{WSR}
  S.~Weinberg,
  %``Precise relations between the spectra of vector and axial vector mesons,''
  Phys.\ Rev.\ Lett.\  {\bf 18}, 507 (1967).

\bibitem{HSWhot}
  M.~Harada, C.~Sasaki and W.~Weise,
  %``Vector-axialvector mixing from a chiral effective field theory at finite temperature,''
  Phys.\ Rev.\ D {\bf 78}, 114003 (2008).

\bibitem{UBW}
  M.~Urban, M.~Buballa and J.~Wambach,
  %``Temperature dependence of rho and alpha(1) meson masses and mixing of vector and axial vector correlators,''
  Phys.\ Rev.\ Lett.\  {\bf 88}, 042002 (2002).

\bibitem{DHholo}
  S.~K.~Domokos and J.~A.~Harvey,
  %``Baryon number-induced Chern-Simons couplings of vector and axial-vector mesons in holographic QCD,''
  Phys.\ Rev.\ Lett.\  {\bf 99}, 141602 (2007).

\bibitem{HSdense}
  M.~Harada and C.~Sasaki,
  %``A Novel spectral broadening from vector--axial-vector mixing in dense matter,''
  Phys.\ Rev.\ C {\bf 80}, 054912 (2009).

\bibitem{Sasaki}
C.~Sasaki,
%``Signatures of chiral symmetry restoration in dilepton production,''
Phys. Lett. B \textbf{801}, 135172 (2020).

\bibitem{kaon}
D.~B.~Kaplan and A.~E.~Nelson,
%``Strange Goings on in Dense Nucleonic Matter,''
Phys. Lett. B \textbf{175}, 57-63 (1986).

\bibitem{KM}
  N.~Kaiser and U.~G.~Meissner,
  %``Generalized hidden symmetry for low-energy hadron physics,''
  Nucl.\ Phys.\ A {\bf 519}, 671 (1990).

\bibitem{BKY}
  M.~Bando, T.~Kugo and K.~Yamawaki,
  %``Nonlinear Realization and Hidden Local Symmetries,''
  Phys.\ Rept.\  {\bf 164}, 217 (1988).

\bibitem{HSghls}
  M.~Harada and C.~Sasaki,
  %``Dropping rho and A(1) meson masses at chiral phase transition in the generalized hidden local symmetry,''
  Phys.\ Rev.\ D {\bf 73}, 036001 (2006).

\bibitem{Marczenko}
  M.~Marczenko, D.~Blaschke, K.~Redlich and C.~Sasaki,
  %``Parity Doubling and the Dense Matter Phase Diagram under Constraints from Multi-Messenger Astronomy,''
  Universe {\bf 5}, no. 8, 180 (2019).

\bibitem{FP}
B.~Friman and H.~J.~Pirner,
%``P wave polarization of the rho meson and the dilepton spectrum in dense matter,''
Nucl. Phys. A \textbf{617}, 496-509 (1997).

\bibitem{RUBW}
R.~Rapp, M.~Urban, M.~Buballa and J.~Wambach,
%``A Microscopic calculation of photoabsorption cross-sections on protons and nuclei,''
Phys. Lett. B \textbf{417}, 1-6 (1998).

\bibitem{giessen}
P.~Muehlich, V.~Shklyar, S.~Leupold, U.~Mosel and M.~Post,
Nucl. Phys. A \textbf{780}, 187-205 (2006).

\bibitem{fastsum}
G.~Aarts, C.~Allton, S.~Hands, B.~J\"ager, C.~Praki and J.~I.~Skullerud,
Phys. Rev. D \textbf{92}, no.1, 014503 (2015);
G.~Aarts, C.~Allton, D.~De Boni, S.~Hands, B.~J\"ager, C.~Praki and J.~I.~Skullerud,
JHEP \textbf{06}, 034 (2017);
G.~Aarts, C.~Allton, D.~De Boni and B.~J\"ager,
Phys. Rev. D \textbf{99}, no.7, 074503 (2019).

\bibitem{Kim:2021xyp}
J.~Kim and S.~H.~Lee,
%``Masses of hadrons in the chiral symmetry restored vacuum,''
Phys. Rev. D \textbf{105}, no.1, 014014 (2022).

\bibitem{Li:1997tz}
G.~Q.~Li, C.~H.~Lee and G.~E.~Brown,
%``Kaon production in heavy ion collisions and maximum mass of neutron stars,''
Phys. Rev. Lett. \textbf{79}, 5214-5217 (1997).

\bibitem{Chung:1998ev}
W.~S.~Chung, C.~M.~Ko and G.~Q.~Li,
%``Seeing phi meson through the dilepton spectra in heavy ion collisions,''
Nucl. Phys. A \textbf{641}, 357-378 (1998).

\bibitem{Oset:2000eg}
E.~Oset and A.~Ramos,
%``Phi decay in nuclei,''
Nucl. Phys. A \textbf{679}, 616-628 (2001).

\bibitem{Glozman:2022lda}
L.~Y.~Glozman, O.~Philipsen and R.~D.~Pisarski,
%``Chiral spin symmetry and the QCD phase diagram,''
[arXiv:2204.05083 [hep-ph]].

\end{thebibliography}
\end{document}